\documentstyle[preprint,aps]{revtex}
\begin{document}

\title{\textbf{Exact solutions of the Schr\"{o}dinger equation in }$D$\textbf{%
-dimensions for the pseudoharmonic potential plus ring-shaped potential }}
\author{Sameer M. Ikhdair\thanks{%
sikhdair@neu.edu.tr} \ and \ Ramazan Sever\thanks{%
sever@metu.edu.tr} \\
%EndAName
{\small \textsl{$^{\ast }$Department of Physics, \ Near East University,
Nicosia, North Cyprus, Mersin-10, Turkey }} \\
{\small \textsl{$^{\dagger }$Department of Physics, Middle East Technical
University, 06531 Ankara, Turkey }}}
\date{\today}
\maketitle

\begin{abstract}
We present analytically the exact energy bound-states solutions of the
Schr\"{o}dinger equation in $D$-dimensions for a pseudoharmonic potential
plus ring-shaped potential of the form $V(r,\theta )=D_{e}\left( \frac{r}{%
r_{e}}-\frac{r_{e}}{r}\right) ^{2}+\frac{\beta \cos ^{2}\theta }{r^{2}\sin
^{2}\theta }$ by means of the conventional Nikiforov-Uvarov method. We also
give a clear recipe of how to obtain an explicit solution to the radial and
angular parts of the wave functions in terms of orthogonal polynomials. The
total energy of the system is different from the pseudoharmonic potential
because of\ the contribution of the angular part. The general results
obtained in this work can be reduced to the standard forms given in the
literature.


PACS\ number: 03.65.-w; 03.65.Fd; 03.65.Ge.
\end{abstract}

% INITIALIZE - DONT CHANGE % %  %

\section{Introduction}

\noindent The solution of the fundamental dynamical equations is an
interesting phenomenon in many fields of physics and chemistry. To obtain
the exact $\ell $-state solutions of the Schr\"{o}dinger equation (SE) are
possible only for a few potentials and hence approximation methods are used
to obtain their solutions [1]. According to the Schr\"{o}dinger formulation
of quantum mechanics, a total wave function provides implicitly all relevant
information about the behaviour of a physical system. Hence if it is exactly
solvable for a given potential, the wave function can describe such a system
completely. Until now, many efforts have been made to solve the stationary
SE with anharmonic potentials in two dimensions $(2D)$ and three dimensions $%
(3D)$ [2-6] with many applications to molecular and chemical physics. The
study of the SE with these potentials provides us with insight into the
physical problem under consideration. However, the study of SE with some of
these potentials in the arbitrary dimensions $D$ is presented in (cf. Ref.
[7] and the references therein). Furthermore, the study of the bound state
processes is also fundamental to understanding molecular spectrum of a
diatomic molecule in quantum mechanics [8].

The Harmonic oscillator [9,10] and H-atom (Coulombic) [9-11] problems have
been thoroughly studied in $D$-dimensional space quantum mechanics for any
angular momentum $\ell .$ These two problems are related together and hence
the resulting second-order differential equation has the normalized
orthogonal polynomial function solution.

On the other hand, the pseudoharmonic potential may be used for the energy
spectrum of linear and non-linear systems [12,13]. It is generally used for
discussion of molecular vibrations. Additionally, this potential possesses
advantages over the harmonic and leads to equally spaced energy levels. One
of the advantages of the pseudoharmonic potential over the harmonic
oscillator is that it can be treated exactly in three as well as in one
dimension. Besides, the Pseudoharmonic and Mie-type potentials [12-14] are
two exactly solvable potentials other than the Coulombic and anharmonic
oscillator [9-11].

An exactly complete bound-state solutions of the $3D$ SE with pseudoharmonic
potential, anharmonic oscillator-like potential with inclusion of a
centrifugal potential barrier does not complicate the solutions, which are
available in closed form. Recently, the solution was also carried out by
using orthogonal polynomial solution method and also by performing a proper
transformation procedures [15]. Besides, the analytical solutions of the $D$%
-dimensional radial SE with some diatomic molecular potentials like
pseudoharmonic [15] and modified Morse or Kratzer-Fues [16] potential are
also solved by selecting a suitable ansatz to the wave function [17].

Chen and Dong [18] found a new ring-shaped (non-central) potential and
obtained the exact solution of the SE for the Coulomb potential plus this
new ring-shaped potential which has possible applications to ring-shaped
organic molecules like cyclic polyenes and benzene. The complete exact
energy bound-state solution and the corresponding wave functions of a class
of non-central potentials [19] have been solved recently by means of the
conventional Nikiforov-Uvarov (NU) method [19-25].

Recently, Cheng and Dai [26], proposed a new potential consisting from the
modified Kratzer's potential [27] plus the new proposed ring-shaped
potential in [18]. They have presented the energy eigenvalues for this
proposed exactly-solvable non-central potential in $3D$-SE throughout the NU
method. The two quantum systems solved by Refs [18,26] are closely relevant
to each other as they deal with a Coulombic field interaction except for
inclusion of a centrifugal potential barrier acts as a repulsive core which
is for any arbitrary angular momentum $\ell $ prevents collapse of the
system in any dimensional space due to this additional perturbation to the
original angular momentum barrier. In a very recent works [28,29], we have
given a clear recipe of how to obtain analytically the exact energy
eigenvalues and the corresponding normalized wave functions of the
Schr\"{o}dinger equation in $D$-dimensions with the proposed modified
Kratzer plus ring-shaped potential [28] and the modified Coulomb plus
ring-shaped potential [29] by means of the conventional NU method [19-29].

The purpose of the paper is to solve the SE for the pseudoharmonic potential
plus this new ring-shaped potential offered in [18]. This new proposed
potential falls among in the class of non-central potential which have
already been solved in our previous work [19]. In spherical coordinates, we
have given this appellation to the non-central potential
\begin{equation}
V(r,\theta )=D_{e}\left( \frac{r}{r_{e}}-\frac{r_{e}}{r}\right) ^{2}+\beta
\frac{ctg^{2}\theta }{r^{2}}=V_{1}(r)+\frac{V_{2}(\theta )}{r^{2}},
\end{equation}
where $D_{e}=kr_{0}^{2}/8$ is the dissociation energy between two atoms in a
solid, $r_{e}$ is the equilibrium intermolecular seperation and $\beta $ is
positive real constant. The potential (1) reduces to the pseudoharmonic
potential in the limiting case of $\beta =0$ [13,15]$.$ The NU method
[19-29] has been used to solve the SE for this new potential (1).

This work is organized as follows: in section \ref{BC}, we shall briefly
introduce the basic concepts of the NU method. Section \ref{ES} is mainly
devoted to the exact solution of the Schr\"{o}dinger equation in $D$%
-dimensions for this quantum system by means of the $\mathrm{NU}$ method.
Finally, the relevant results are discussed in section \ref{C}.

\section{Basic Concepts of the Method}

\label{BC}The NU method is based on reducing the second-order differential
equation to a generalized equation of hypergeometric type [20]. In this
sense, the Schr\"{o}dinger equation, after employing an appropriate
coordinate transformation $s=s(r),$ transforms to the following form:
\begin{equation}
\psi _{n}^{\prime \prime }(s)+\frac{\widetilde{\tau }(s)}{\sigma (s)}\psi
_{n}^{\prime }(s)+\frac{\widetilde{\sigma }(s)}{\sigma ^{2}(s)}\psi
_{n}(s)=0,
\end{equation}
where $\sigma (s)$ and $\widetilde{\sigma }(s)$ are polynomials, at most of
second-degree, and $\widetilde{\tau }(s)$ is a first-degree polynomial.
Using a wave function, $\psi _{n}(s),$ of \ the simple ansatz:

\begin{equation}
\psi _{n}(s)=\phi _{n}(s)y_{n}(s),
\end{equation}
reduces (2) into an equation of a hypergeometric type

\begin{equation}
\sigma (s)y_{n}^{\prime \prime }(s)+\tau (s)y_{n}^{\prime }(s)+\lambda
y_{n}(s)=0,
\end{equation}
where

\begin{equation}
\sigma (s)=\pi (s)\frac{\phi (s)}{\phi ^{\prime }(s)},
\end{equation}

\begin{equation}
\tau (s)=\widetilde{\tau }(s)+2\pi (s),\text{ }\tau ^{\prime }(s)<0,
\end{equation}
and $\lambda $ is a parameter defined as
\begin{equation}
\lambda =\lambda _{n}=-n\tau ^{\prime }(s)-\frac{n\left( n-1\right) }{2}%
\sigma ^{\prime \prime }(s),\text{ \ \ \ \ \ \ }n=0,1,2,....
\end{equation}
The polynomial $\tau (s)$ with the parameter $s$ and prime factors show the
differentials at first degree be negative. It is worthwhile to note that $%
\lambda $ or $\lambda _{n}$ are obtained from a particular solution of the
form $y(s)=y_{n}(s)$ which is a polynomial of degree $n.$ Further, the other
part $y_{n}(s)$ of the wave function (3) is the hypergeometric-type function
whose polynomial solutions are given by Rodrigues relation

\begin{equation}
y_{n}(s)=\frac{B_{n}}{\rho (s)}\frac{d^{n}}{ds^{n}}\left[ \sigma ^{n}(s)\rho
(s)\right] ,
\end{equation}
where $B_{n}$ is the normalization constant and the weight function $\rho
(s) $ must satisfy the condition [20]

\begin{equation}
\frac{d}{ds}w(s)=\frac{\tau (s)}{\sigma (s)}w(s),\text{ }w(s)=\sigma (s)\rho
(s).
\end{equation}
The function $\pi $ and the parameter $\lambda $ are defined as

\begin{equation}
\pi (s)=\frac{\sigma ^{\prime }(s)-\widetilde{\tau }(s)}{2}\pm \sqrt{\left(
\frac{\sigma ^{\prime }(s)-\widetilde{\tau }(s)}{2}\right) ^{2}-\widetilde{%
\sigma }(s)+k\sigma (s)},
\end{equation}
\begin{equation}
\lambda =k+\pi ^{\prime }(s).
\end{equation}
In principle, since $\pi (s)$ has to be a polynomial of degree at most one,
the expression under the square root sign in (10) can be arranged to be the
square of a polynomial of first degree [20]. This is possible only if its
discriminant is zero. In this case, an equation for $k$ is obtained. After
solving this equation, the obtained values of $k$ are substituted in (10).
In addition, by comparing equations (7) and (11), we obtain the energy
eigenvalues.

\section{Exact solutions of the quantum system with the NU method}

\label{ES}

\subsection{Seperating variables of the Schr\"{o}dinger equation}

The proposed potential (1) can be simply rewritten in the form of isotropic
harmonic oscillator plus inverse quadratic plus ring-shaped potential as
\begin{equation}
V(r,\theta )=ar^{2}+\frac{b}{r^{2}}+\beta \frac{\cos ^{2}\theta }{r^{2}\sin
^{2}\theta }+c,\text{ }\beta >0
\end{equation}
where $a=D_{e}r_{e}^{-2},$ $b=D_{e}r_{e}^{2}$ and $c=-2D_{e}.$

The potential in (12) solved for the limiting case of $\beta =0$ by using
the orthogonal polynomial solution method [15] and by the Rydberg-Klein-Rees
(RKR) procedures [13]. In fact the energy spectrum for the potential in (12)
can be obtained directly by considering it as one case of the general
non-central seperable potentials discussed previously in [19].

Our aim is to derive analytically the exact energy spectrum for a moving
particle in the presence of a potential (12) in a very simple way. The $D$%
-dimensional space SE, in spherical polar coordinates, for the potential
(12) is [28-30]

\[
-\frac{\hbar ^{2}}{2\mu }\left\{ \frac{1}{r^{D-1}}\frac{\partial }{\partial r%
}\left( r^{D-1}\frac{\partial }{\partial r}\right) +\frac{1}{r^{2}}\left[
\frac{1}{\sin \theta }\frac{\partial }{\partial \theta }\left( \sin \theta
\frac{\partial }{\partial \theta }\right) \right. \right.
\]
\begin{equation}
+\left. \left. \frac{1}{\sin ^{2}\theta }\frac{\partial ^{2}}{\partial
\varphi ^{2}}-\frac{2\mu \beta }{\hbar ^{2}}\frac{\cos ^{2}\theta }{\sin
^{2}\theta }\right] +\frac{2\mu }{\hbar ^{2}}\left( E-ar^{2}-\frac{b}{r^{2}}%
-c\right) \right\} \psi (r,\theta ,\varphi )=0,
\end{equation}
where $\mu =\frac{m_{1}m_{2}}{m_{1}+m_{2}}$ being the reduced mass of the
two particles and $\psi (r,\theta ,\varphi )$ being the total wave function
separated as follows

\begin{equation}
\psi _{n\ell m}(r,\theta ,\varphi )=R(r)Y_{\ell }^{m}(\theta ,\varphi ),%
\text{ }R(r)=r^{-(D-1)/2}g(r),\text{ }Y_{\ell }^{m}(\theta ,\varphi
)=H(\theta )\Phi (\varphi ).
\end{equation}
On substituting equation (14) into (13) leads to a set of second-order
differential equations:
\begin{equation}
\frac{d^{2}\Phi (\varphi )}{d\varphi ^{2}}+m^{2}\Phi (\varphi )=0,
\end{equation}

\begin{equation}
\left[ \frac{1}{\sin \theta }\frac{d}{d\theta }\left( \sin \theta \frac{d}{%
d\theta }\right) -\frac{m^{2}}{\sin ^{2}\theta }-\frac{2\mu \beta }{\hbar
^{2}}\frac{\cos ^{2}\theta }{\sin ^{2}\theta }+\ell (\ell +D-2)\right]
H(\theta )=0,
\end{equation}
\begin{equation}
\left[ \frac{1}{r^{D-1}}\frac{d}{dr}\left( r^{D-1}\frac{d}{dr}\right) -\frac{%
\ell (\ell +D-2)}{r^{2}}\right] R(r)+\frac{2\mu }{\hbar ^{2}}\left[ E-ar^{2}-%
\frac{b}{r^{2}}-c\right] R(r)=0.
\end{equation}
The solution in (15) is periodic and must satisfy the period boundary
condition $\Phi (\varphi +2\pi )=\Phi (\varphi )$ from which we obtain
[9,10]
\begin{equation}
\Phi _{m}(\varphi )=\frac{1}{\sqrt{2\pi }}\exp (\pm im\varphi ),\text{ \ }%
m=0,1,2,.....
\end{equation}
Further, equation (16) representing the angular wave equation takes the
simple form
\begin{equation}
\frac{d^{2}H(\theta )}{d\theta ^{2}}+\frac{\cos \theta }{\sin \theta }\frac{%
dH(\theta )}{d\theta }+\left[ \ell (\ell +D-2)-\frac{m^{2}+(2\mu \beta
/\hbar ^{2})\cos ^{2}\theta }{\sin ^{2}\theta }\right] H(\theta )=0,
\end{equation}
which will be solved in the following subsection.

\subsection{The solutions of the angular equation}

In order to apply NU method, we introduce a new variable $s=\cos \theta ,$
Eq. (19) is then rearranged as the universal associated-Legendre
differential equation [26,28,29]

\begin{equation}
\frac{d^{2}H(s)}{ds^{2}}-\frac{2s}{1-s^{2}}\frac{dH(s)}{ds}+\frac{\nu
^{\prime }(1-s^{2})-m^{\prime }{}^{2}}{\sin ^{2}\theta }H(\theta )=0,
\end{equation}
where

\begin{equation}
\nu ^{\prime }=\ell ^{\prime }(\ell ^{\prime }+D-2)=\ell (\ell +D-2)+2\mu
\beta /\hbar ^{2}\text{ \ \ and \ \ \ }m^{\prime }{}=\sqrt{m^{2}+2\mu \beta
/\hbar ^{2}}.
\end{equation}
The solution of this equation has already been solved by the NU method in
[26,28,29]. However, the aim in this subsection is to solve with different
parameters resulting from the $D$-space-dimensions of Schr\"{o}dinger
equation. Upon letting $D=3,$ we can readily obtain the standard case given
in [26]. By comparing Eqs. (20) and (2), the corresponding polynomials are
obtained. In terms of the variable $s,$ we have
\begin{equation}
\widetilde{\tau }(s)=-2s,\text{ \ \ \ }\sigma (s)=1-s^{2},\text{ \ \ }%
\widetilde{\sigma }(s)=-\nu ^{\prime }s^{2}+\nu ^{\prime }-m^{\prime }{}^{2}.
\end{equation}
Inserting the above expressions into equation (10) and taking $\sigma
^{\prime }(s)=-2s$, one obtains the following function:

\begin{equation}
\pi (s)=\pm \sqrt{(\nu ^{\prime }-k)s^{2}+k-\nu ^{\prime }+m^{\prime }{}^{2}}%
.
\end{equation}
Following the method, the polynomial $\pi (s)$ is found to have the
following four possible values:
\begin{equation}
\pi (s)=\left\{
\begin{array}{cc}
m^{\prime }s & \text{\ for }k_{1}=\nu ^{\prime }-m^{\prime }{}^{2}, \\
-m^{\prime }s & \text{\ for }k_{1}=\nu ^{\prime }-m^{\prime }{}^{2}, \\
m^{\prime } & \text{\ for }k_{2}=\nu ^{\prime }, \\
-m^{\prime } & \text{\ for }k_{2}=\nu ^{\prime }.
\end{array}
\right.
\end{equation}
Imposing the condition $\tau ^{\prime }(s)<0,$ for equation (6), one selects
the following physically valid solutions:

\begin{equation}
k_{1}=\nu ^{\prime }-m^{\prime }{}^{2}\text{ \ \ and \ \ }\pi (s)=-m^{\prime
}s,
\end{equation}
which yields form equation (6)
\begin{equation}
\tau (s)=-2(1+m^{\prime })s.
\end{equation}
Using Eqs (7) and (11), the following expressions for $\lambda $ are
obtained, respectively,

\begin{equation}
\lambda =\lambda _{n}=2n(1+m^{\prime })+n(n-1),
\end{equation}
\begin{equation}
\lambda =\nu ^{\prime }-m^{\prime }{}(1+m^{\prime }).
\end{equation}
We compare Eqs (27) and (28) and from the definition $\nu ^{\prime }=\ell
^{\prime }(\ell ^{\prime }+D-2),$ the new angular momentum\ $\ell ^{\prime }$
values are obtained as

\begin{equation}
\ell ^{\prime }=-\frac{(D-2)}{2}+\frac{1}{2}\sqrt{(D-2)^{2}+4(n+m^{\prime
})(n+m^{\prime }+1)},
\end{equation}
which can be easily reduced to the simple form

\begin{equation}
\ell ^{\prime }=n+m^{\prime },
\end{equation}
in $3D$ [26]. Using Eqs (3)-(5) and (8)-(9), the wave function can be
written as,

\begin{equation}
H_{m^{\prime }}(\theta )=\sqrt{\frac{(2\ell ^{\prime }+1)(\ell ^{\prime
}-m^{\prime })!}{2(\ell ^{\prime }+m^{\prime })!}}\sin ^{m^{\prime }}(\theta
)P_{n}^{(m^{\prime },m^{\prime })}(\cos \theta ),
\end{equation}
where
\begin{equation}
n=-\frac{(1+2m^{\prime })}{2}+\frac{1}{2}\sqrt{(2\ell ^{\prime
}+1)^{2}+4\ell ^{\prime }(D-3)},
\end{equation}
with $\ell ^{\prime }$ is given explicitly in (29). \

Hence, we are left to solve Eq. (17). It transforms, after lengthy but
straightforward algebra, into the following simple form [30]:
\begin{equation}
\frac{d^{2}g(r)}{dr^{2}}+\left[ \frac{2\mu }{\hbar ^{2}}(E-c)-\frac{2\mu a}{%
\hbar ^{2}}r^{2}-\frac{\widetilde{\nu }+(2\mu b/\hbar ^{2})}{r^{2}}\right]
g(r)=0,
\end{equation}
where
\begin{equation}
\widetilde{\nu }=\frac{1}{4}(M-1)(M-3),\text{ }M=D+2\ell .
\end{equation}
Obviously, the two particles in Eq. (33) interacting via anharmonic
oscillator potential plus inclusion of a centrifugal potential acts as a
repulsive core which for any arbitrary $\ell $ prevents collapse of the
system in any space dimension due to this additional centrifugal potential
barrier. Hence, Eq. (33) has to be solved by using the NU method in the next
subsection.

\subsection{The solutions of the radial equation}

The aim of this subsection is to solve the problem with a different radial
separation function $g(r)$ in any arbitrary dimensions. We now study the
bound-states (real) solution $E>c$ of Eq. (33). By employing the suitable
transformation, $s=r^{2},$ and letting

\begin{equation}
\varepsilon =\sqrt{\frac{2\mu }{\hbar ^{2}}(E-c)},\text{ }\alpha =\frac{2\mu
a}{\hbar ^{2}},\text{ }\gamma =\widetilde{\nu }+\frac{2\mu b}{\hbar ^{2}},
\end{equation}
one can transforms Eq. (33) into the following form:

\begin{equation}
\frac{d^{2}g(s)}{ds^{2}}+\frac{1}{2s}\frac{dg(s)}{ds}+\frac{1}{\left(
2s\right) ^{2}}\left[ \varepsilon ^{2}s-\alpha s^{2}-\gamma \right] g(s)=0.
\end{equation}
To apply the conventional NU-method, Eq. (36) is compared with (2), the
corresponding polynomials are obtained

\begin{equation}
\widetilde{\tau }(s)=1,\text{ \ \ \ }\sigma (s)=2s,\text{ \ \ }\widetilde{%
\sigma }(s)=\varepsilon ^{2}s-\alpha s^{2}-\gamma .
\end{equation}
The polynomial $\pi (s)$ in Eq. (10) can be found by substituting Eq. (37)
and taking $\sigma ^{\prime }(s)=2.$ Hence, the polynomial $\pi (s)$ is

\begin{equation}
\pi (s)=\frac{1}{2}\pm \frac{1}{2}\sqrt{4\alpha s^{2}+4(2k-\varepsilon
^{2})s+4\gamma +1}.
\end{equation}
According to this conventional method, the form of the expression under the
square root in Eq. (38) is to be set equal to zero and solved for the two
roots of $k$ which can be readily obtained as

\begin{equation}
k=\frac{\varepsilon ^{2}}{2}\pm \frac{1}{2}\sqrt{\alpha (4\gamma +1)}.
\end{equation}
In view of that, we arrive at the following four possible functions of $\pi
(s):$%
\begin{equation}
\pi (s)=\left\{
\begin{array}{cc}
\frac{1}{2}+\frac{1}{2}\left[ 2\sqrt{\alpha }s+\sqrt{4\gamma +1}\right] &
\text{\ for }k_{1}=\frac{\varepsilon ^{2}}{2}+\frac{1}{2}\sqrt{\alpha
(4\gamma +1)}, \\
\frac{1}{2}-\frac{1}{2}\left[ 2\sqrt{\alpha }s+\sqrt{4\gamma +1}\right] &
\text{\ for }k_{1}=\frac{\varepsilon ^{2}}{2}+\frac{1}{2}\sqrt{\alpha
(4\gamma +1)}, \\
\frac{1}{2}+\frac{1}{2}\left[ 2\sqrt{\alpha }s-\sqrt{4\gamma +1}\right] &
\text{\ for }k_{2}=\frac{\varepsilon ^{2}}{2}-\frac{1}{2}\sqrt{\alpha
(4\gamma +1)}, \\
\frac{1}{2}-\frac{1}{2}\left[ 2\sqrt{\alpha }s-\sqrt{4\gamma +1}\right] &
\text{\ for }k_{2}=\frac{\varepsilon ^{2}}{2}-\frac{1}{2}\sqrt{\alpha
(4\gamma +1)}.
\end{array}
\right.
\end{equation}
The correct value of $\pi (s)$ is chosen such that the function $\tau (s)$
given by Eq. (6) will have negative derivative [20]. So we can select the
physically valid solutions to be

\begin{equation}
k=\frac{\varepsilon ^{2}}{2}-\frac{1}{2}\sqrt{\alpha (4\gamma +1)}\text{ \ \
and \ \ }\pi (s)=\frac{1+\sqrt{4\gamma +1}}{2}-\sqrt{\alpha }s,
\end{equation}
which give
\begin{equation}
\tau (s)=2+\sqrt{4\gamma +1}-2\sqrt{\alpha }s,\text{ }\tau ^{\prime }(s)=-2%
\sqrt{\alpha }<0.
\end{equation}
Using Eqs (7) and (11), the following expressions for $\lambda $ are
obtained, respectively,

\begin{equation}
\lambda =\lambda _{n}=2N\sqrt{\alpha },\text{ }N=0,1,2,...,
\end{equation}
\begin{equation}
\lambda =\frac{\varepsilon ^{2}}{2}-\frac{1}{2}\sqrt{\alpha (4\gamma +1)}-%
\sqrt{\alpha }.
\end{equation}
Hence, the energy eigenvalues are:

\begin{equation}
E_{N}=c+\sqrt{\frac{\hbar ^{2}a}{2\mu }}\left( 4N+2+\sqrt{(M-1)(M-3)+8\mu
b/\hbar ^{2}+1}\right) .
\end{equation}
Substituting

\begin{equation}
(M-1)(M-3)=4\widetilde{\nu }=(D-2)^{2}+4\ell ^{\prime }(\ell ^{\prime
}+D-2)-8\mu \beta /\hbar ^{2}-1,
\end{equation}
with $\ell ^{\prime }$ defined in (29), Eq. (45) transforms into the form
\begin{equation}
E_{N}=c+\sqrt{\frac{\hbar ^{2}a}{2\mu }}\left( 4N+2+\sqrt{(D-2)^{2}+4\ell
^{\prime }(\ell ^{\prime }+D-2)+8\mu (b-\beta )/\hbar ^{2}}\right) ,\text{ }
\end{equation}
where $N,\ell ^{\prime }=0,1,2,...$ Equation (47) contains the contributions
coming from the angular-dependent part of the SE for the pseudoharmonic
potential plus ring-shaped potential as well.

(i) For $3D$-pseudoharmonic potential plus ring-shaped potential case, we
use transformation of parameters $a=D_{e}r_{e}^{-2},$ $b=D_{e}r_{e}^{2}$ and
$c=-2D_{e}$ in Eq. (47):
\begin{equation}
E_{N}=-2D_{e}+\sqrt{\frac{2\hbar ^{2}D_{e}}{\mu r_{e}^{2}}}\left( 2N+1+\sqrt{%
(n+m^{\prime }+1/2)^{2}+2\mu (D_{e}r_{e}^{2}-\beta )/\hbar ^{2}}\right) ,
\end{equation}
where $N,n,m^{\prime }=0,1,2,..$ and $m^{\prime }$ is defined in Eq. (21).

(ii) For $3D$-pseudoharmonic potential case, we use the transformation of
the parameters $a=D_{e}r_{e}^{-2},$ $b=D_{e}r_{e}^{2},$ $c=-2D_{e}$ in Eq.
(47):
\begin{equation}
E_{N}=-2D_{e}+\sqrt{\frac{2\hbar ^{2}}{\mu r_{e}^{2}}}\left( 2N+1+\sqrt{%
\left( \ell +\frac{1}{2}\right) ^{2}+2\mu D_{e}r_{e}^{2}/\hbar ^{2}}\right) ,%
\text{ }N,\ell =0,1,2,...
\end{equation}
where $\ell =n+m$ and it is found to be consistent with Ref. [15].

In what follows, we attempt to find the radial wavefunctions for this
potential. Using $\tau (s),$ $\pi (s)$ and $\sigma (s)$ in Eqs (5) and (9),
we find the first part of the wave function
\begin{equation}
\phi (s)=\exp \left( -\frac{\sqrt{\alpha }}{2}s\right) s^{(1+\sqrt{4\gamma +1%
})/4},
\end{equation}
and the weight function
\begin{equation}
\rho (s)=\exp \left( -\sqrt{\alpha }s\right) s^{\sqrt{4\gamma +1}/2}
\end{equation}
which is useful for finding the second part of the wave function. Besides,
we substitute Eq. (51) into the Rodrigues relation in Eq. (8) and obtain one
of the wave functions in the form
\begin{equation}
y_{n}(s)=B_{n}\exp \left( \sqrt{\alpha }s\right) s^{-\sqrt{4\gamma +1}/2}%
\frac{d^{N}}{ds^{N}}\left( s^{N+\sqrt{4\gamma +1}/2}\exp \left( -\sqrt{%
\alpha }s\right) \right) ,
\end{equation}
where $B_{n}$ is a normalization constant. Hence, the wave function $g(s)$
can be written in the form of the generalized Laguerre polynomials as

\begin{equation}
g(r)=C_{N,L}\exp \left( -\frac{\sqrt{\alpha }}{2}s\right) s^{(1+\sqrt{%
4\gamma +1})/4}L_{N}^{(L+\frac{1}{2})}(\sqrt{\alpha }s),
\end{equation}
where

\begin{equation}
L=\frac{1}{2}\left[ \sqrt{(D-2)^{2}+4\ell ^{\prime }(\ell ^{\prime
}+D-2)+8\mu (b-\beta )/\hbar ^{2}}-1\right] .
\end{equation}
Finally, the radial wave functions of the Schr\"{o}dinger equation are
obtained
\begin{equation}
R(r)=C_{N,L}r^{L-(D-3)/2}\exp \left( -\frac{\sqrt{\alpha }}{2}r^{2}\right)
L_{N}^{(L+\frac{1}{2})}(\sqrt{\alpha }r^{2}),
\end{equation}
where $C_{N,L}$ is the normalization constant to be determined below. Using
the normalization condition, $\int\limits_{0}^{\infty }R^{2}(r)r^{D-1}dr=1,$
and the orthogonality relation of the generalized Laguerre polynomials, $%
\int\limits_{0}^{\infty }z^{\eta +1}e^{-z}\left[ L_{n}^{\eta }(z)\right]
^{2}dz=\frac{(2n+\eta +1)(n+\eta )!}{n!},$ we have

\begin{equation}
C_{N,L}=\sqrt{\frac{2\left( \sqrt{2\mu D_{e}}/\hbar r_{e}\right) ^{L+3/2}N!}{%
\Gamma (L+N+3/2)}}.
\end{equation}
Therefore, we may express the normalized total wave functions as

\[
\psi (r,\theta ,\varphi )=\sqrt{\frac{\left( \sqrt{2\mu D_{e}}/\hbar
r_{e}\right) ^{L+3/2}(2\ell ^{\prime }+1)(\ell ^{\prime }-m^{\prime })!N!}{%
\pi (\ell ^{\prime }+m^{\prime })!\Gamma (N+L+3/2)}}r^{L-(D-3)/2}
\]
\begin{equation}
\times \exp \left( -\sqrt{\frac{\mu D_{e}}{2\hbar ^{2}r_{e}^{2}}}%
r^{2}\right) L_{N}^{(L+1/2)}\left( \sqrt{\frac{2\mu D_{e}}{\hbar
^{2}r_{e}^{2}}}r^{2}\right) \sin ^{m^{\prime }}(\theta )P_{n}^{(m^{\prime
},m^{\prime })}(\cos \theta )\exp (\pm im\varphi ).
\end{equation}
On the other hand, the wave functions in (57) can be reduced to their
standard forms in [15] as one sets $\beta =0.$ Therefore, we finally have

\[
\psi (r,\theta ,\varphi )=\sqrt{\frac{\left( \sqrt{2\mu D_{e}}/\hbar
r_{e}\right) ^{L+3/2}(2\ell +1)(\ell -m)!N!}{\pi (\ell +m)!(N+L+1/2)!}}%
r^{L-(D-3)/2}
\]
\begin{equation}
\times \exp \left( -\sqrt{\frac{\mu D_{e}}{2\hbar ^{2}r_{e}^{2}}}%
r^{2}\right) L_{N}^{(L+1/2)}\left( \sqrt{\frac{2\mu D_{e}}{\hbar
^{2}r_{e}^{2}}}r^{2}\right) \sin ^{m}(\theta )P_{n}^{(m,m)}(\cos \theta
)\exp (\pm im\varphi ).
\end{equation}
where
\begin{equation}
L+\frac{1}{2}=\sqrt{(\ell +1/2)^{2}+2\mu D_{e}r_{e}^{2}/\hbar ^{2}},\text{ }%
\ell =n+m.
\end{equation}
By choosing appropriate values of the parameters $D_{e}$ and $r_{e},$ the
desired exact energy spectrum and the corresponding wave functions can be
calculated for every case.

\section{ Conclusions}

\label{C}We have easily obtained the exact bound state solutions of the $D$
dimensional radial SE for a diatomic molecule with the pseudoharmonic
potential plus ring-shaped potential by means of the conventional NU method
for any $N\ell ^{\prime }m^{\prime }$. The NU method can be applied
systematically to both radial and angular parts of the wave function.
Further, the wave functions are expressed in terms of special orthogonal
functions such as Laguerre and Jacobi polynomials [31]. The presented
procedure in this study is systematical and efficient in finding the exact
energy spectra and corresponding wave functions of the Schr\"{o}dinger
equation for various diatomic molecules. This new proposed potential can be
reduced to the standard pseudoharmonic potential in the $3D,$ which appears
to describe the molecular vibrations quite well, by doing the following
transformations: $m^{\prime }\rightarrow m,$ $\ell ^{\prime }\rightarrow
\ell =n+m.$ We point out that this method is simple and promising in
producing the exact bound state solution for energy states and wave
functions for further anharmonic oscillator-type potential plus new
ring-shaped potential.

\section{Acknowledgments}

This research was partially supported by the Scientific and Technological
Research Council of Turkey.

\end{document}